\documentclass[12pt,preprint]{aastex} 

\shorttitle{\sc The Missing Link in Star Cluster Evolution}

\begin{document}

\title{The Missing Link in Star Cluster Evolution}

\author{Richard de Grijs,}
\affil{Institute of Astronomy, University of Cambridge, Madingley Road,
Cambridge CB3 0HA, UK}
\author{Nate Bastian, and Henny J.G.L.M. Lamers}
\affil{Astronomical Institute, Utrecht University Princetonplein 5,
3584 CC Utrecht, The Netherlands}

\begin{abstract} 
The currently most popular models for the dynamical evolution of star
clusters predict that the power-law Cluster Luminosity Functions (CLFs)
of young star cluster systems will be transformed rapidly into the
universal Gaussian CLFs of old Milky Way-type ``globular'' cluster
systems.  Here, we provide the first evidence for a turn-over in
the intermediate-age, approximately 1 Gyr-old CLF in the center of the
nearby starburst galaxy M82, which very closely matches the universal
CLFs of old Milky Way-type globular cluster systems.  This provides an
important test of both cluster disruption theories and hierarchical
galaxy formation models.  It also lends strong support to the scenario
that these young cluster systems may eventually evolve into old Milky
Way-type globular cluster systems.  M82's proximity, its shortest known
cluster disruption time-scale of any galaxy, and its well-defined peak
of cluster formation make it an ideal candidate to probe the evolution
of its star cluster system to fainter luminosities, and thus lower
masses, than has been possible for any galaxy before.
\end{abstract}


\section{The Cluster Luminosity Function}

For old globular cluster systems, with ages in excess of $\sim 10^{10}$
yr, the shape of the cluster luminosity function (CLF) is
well-established: it is roughly Gaussian, with an apparently universal
peak luminosity, $M_V^0 \simeq -7.4$ mag, and a Gaussian FWHM $\sim 3$
mag (e.g., Whitmore et al.  1995, Harris 1996, 2001, Ashman \& Zepf
1998, Harris et al.  1998).  On the other hand, the well-studied young,
$\lesssim 2$ Gyr-old star cluster population in the Large Magellanic
Cloud displays a power-law CLF of the form $N(L) {\rm d} L \propto
L^{\alpha} {\rm d} L$, where $L$ is the cluster luminosity and $-2.0
\lesssim \alpha \lesssim -1.5$ (e.g., Elson \& Fall 1985, Elmegreen \&
Efremov 1997). 

{\sl Hubble Space Telescope (HST)} observations are continuing to
provide an increasing number of luminosity distributions for additional
samples of young and intermediate-age compact star clusters in more
distant galaxies (e.g., Whitmore \& Schweizer 1995, Schweizer et al. 
1996, Miller et al.  1997, Zepf et al.  1999, de Grijs, O'Connell \&
Gallagher 2001, Whitmore et al.  2002).  Although a large number of
studies have attempted to detect a turn-over in young or
intermediate-age CLFs, the shapes of such young and intermediate-age
CLFs have thus far been consistent with power laws down to the
observational completeness thresholds. 

Based on the observational evidence discussed above, the currently most
popular globular cluster formation models suggest that the distribution
of the initial cluster luminosities and, therefore, the corresponding
distribution of their initial masses is closely approximated by a power
law (Harris \& Pudritz 1994, McLaughlin \& Pudritz 1996, Elmegreen \&
Efremov 1997, Ashman \& Zepf 2001) of the form d$N(M) {\rm d}M \propto
M^{\alpha} {\rm d}M$, where $-2.0 \lesssim \alpha \lesssim -1.5$. 
Several processes are proposed for the transformation of the CLF of
young star cluster systems into the Gaussian or log-normal distributions
characteristic of old globular cluster-type distributions.  These
include the preferential depletion of low-mass clusters both by
evaporation due to two-body relaxation and by tidal interactions with
the gravitational field of their host galaxy, and the preferential
disruption of high-mass clusters by dynamical friction.  However, the
relative importance of these disruption processes is still controversial
(see, e.g., Vesperini [2000, 2001], and Bromm \& Clarke [2002] for
models that do not implicitly assume a power-law initial mass
distribution).  If the age range within a given cluster system is a
significant fraction of the system's mean age, fading due to aging of
the non-coeval cluster population also severely affects the shape of its
CLF: the observed CLF will de dominated by the younger clusters, because
a fraction of the older clusters in the system will have faded to below
the observational detection limit, thus introducing further
incompleteness effects. 

The detection of a characteristic turn-over luminosity in young or
intermediate-age CLFs would lend strong support to the popular (but thus
far only speculative) scenario that the young star clusters observed in
mergers are the analogues to the ubiquitous old globular clusters at
younger ages. 

\section{The intermediate-age star cluster system in M82}

We now consider the intermediate-age star cluster system in the
disturbed, late-type galaxy M82 in this context. M82, the nearest and
best-studied starburst galaxy, has undergone multiple starburst episodes
over the past $\sim 1-2$ Gyr. These were likely triggered by tidal
interactions with M81 and/or other members of the M81/M82/NGC 3077 group
(O'Connell \& Mangano 1978, Telesco 1988, Rieke et al. 1993, Yun, Ho \&
Lo 1994, O'Connell et al. 1995, de Grijs et al. 2001, 2003). 

In a recent study in which we focused on the fossil starburst site near
the center of M82, region B, we found a population of $\sim 110$
gravitationally bound evolved compact star clusters (de Grijs et al. 
2001). Their properties appear to be consistent with the conclusion
that they are evolved counterparts of the young compact star clusters
detected in the galaxy's active core (O'Connell et al. 1995). We
estimated ages for the M82 B cluster population from $\sim 30$ Myr to
over 10 Gyr, with a peak at 1.0 Gyr (de Grijs et al. 2001, 2003), based
on a comparison of broad-band optical and near-infrared {\sl HST}
colors with the Bruzual \& Charlot (2000; BC00) stellar evolutionary
synthesis models. 

\begin{figure}
\plotone{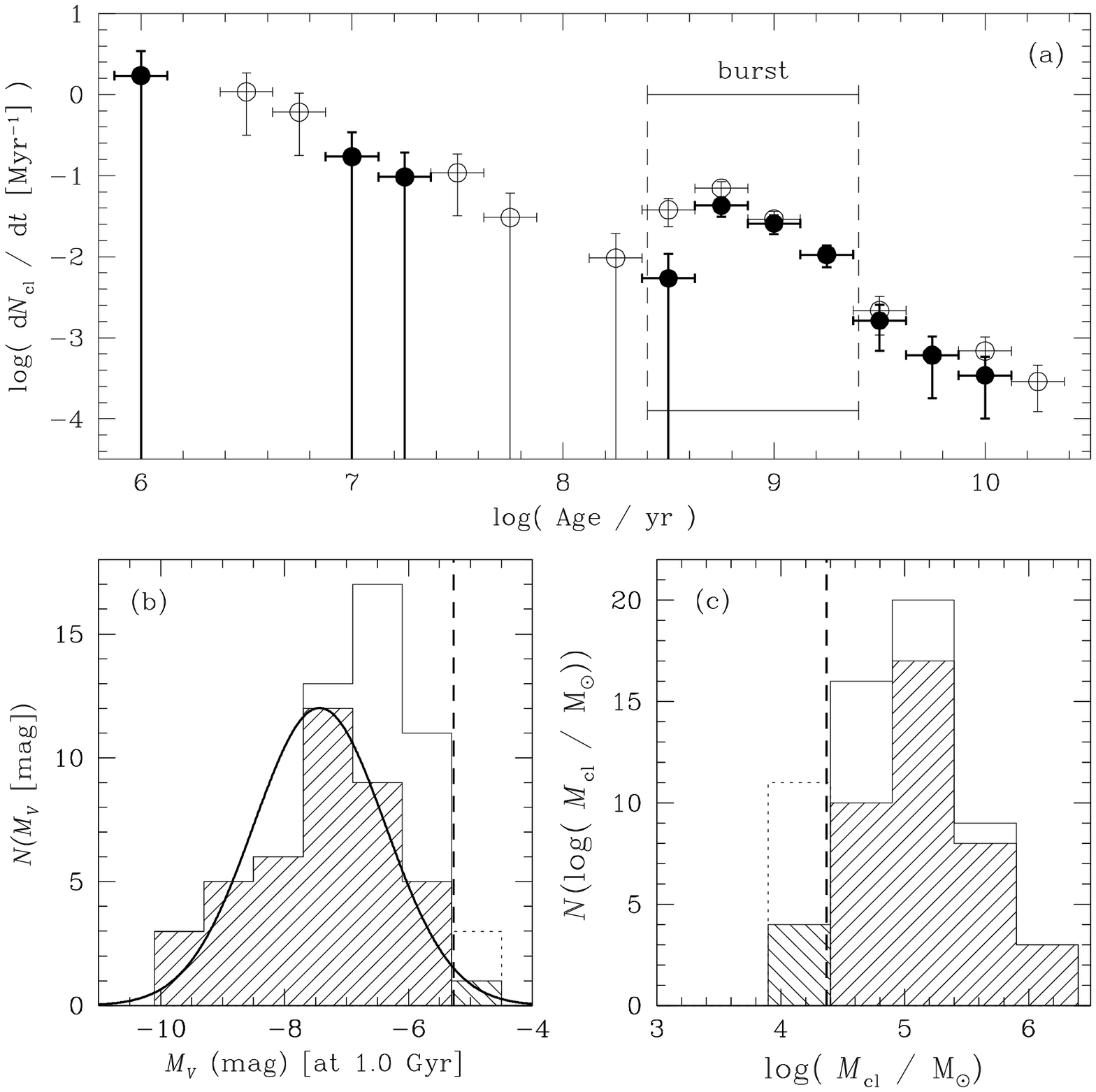}
\figcaption[f1.eps]{\label{m82clf.fig} {\it (a)} The cluster
formation rate (in number of clusters per Myr) as a function of age. 
Open circles: full sample; filled circles: clusters with well-determined
ages (de Grijs et al.  2003).  The age range dominated by the burst of
cluster formation is indicated.  {\it (b)} CLF of the clusters formed in
the burst of cluster formation, $8.4 \le \log ({\rm Age/yr}) \le 9.4$. 
The shaded histograms correspond to the clusters with well-determined
ages; the open histograms represent the entire cluster sample in this
age range, as above.  The Gaussian curve is the best fit to the shaded
distribution.  Finally, the vertical dashed line is our selection limit. 
{\it (c)} Mass distribution of the clusters formed in the burst of
cluster formation; the coding is as in Fig.  \ref{m82clf.fig}b.  The
expected effect of the $A_V \lesssim 0.2$ mag extinction (de Grijs et
al.  2003) for the clusters with well-determined ages is a shift in mass
towards higher masses of $\Delta \log(M_{\rm cl}/M_\odot) \lesssim
0.08$, which implies that the observed turnover is not a spurious effect
due to varying extinction.}
\end{figure}

These results suggest steady, continuing cluster formation in M82 B at a
very modest rate at early times ($> 2$ Gyr ago) followed by a
concentrated formation episode lasting from 500--1500 Myr ago and a
subsequent suppression or decline of cluster formation (de Grijs et al. 
2001, 2003, Parmentier, de Grijs \& Gilmore 2003).  This finite burst of
cluster formation makes the star cluster system in M82 B a good
candidate to address the evolution of the CLF, because it provides a
large sample of approximately coeval clusters (see Fig. 
\ref{m82clf.fig}a).  Moreover, because of the proximity of M82, we have
been able to probe the intermediate-age cluster population in M82 B to
fainter absolute magnitudes, and thus lower masses, than has been
possible before in other, more distant galaxies. 

Figure \ref{m82clf.fig}a shows the star cluster formation rate in M82 B
(de Grijs et al.  2003).  We have highlighted the enhanced cluster
formation episode, $8.4 \lesssim \log ({\rm Age / yr}) \lesssim 9.4$. 
Uncertainties in the age determinations 
may
have broadened the peak, so that the actual duration of the burst of
cluster formation may have been shorter (de Grijs et al.  2001, 2003). 

For the proper interpretation of the M82 B CLF, we need to correct the
individual cluster luminosities for their range in ages (Meurer 1995,
Fritze--v.  Alvensleben 1999, de Grijs et al.  2001, 2003).  Using the
BC00 models, we have corrected the present-day absolute magnitudes of
the clusters formed in the burst of cluster formation to those at a
common, fiducial age of 1.0 Gyr.  The results are shown in Fig. 
\ref{m82clf.fig}b for both the subsample of 42 clusters with
well-defined ages (de Grijs et al.  2003), and for the full sample of 58
clusters formed in the burst (open histogram).  The corresponding mass
distributions, obtained from the application of the BC00
mass--luminosity relation to the cluster luminosities, are shown in Fig. 
\ref{m82clf.fig}c.  In both Figs.  \ref{m82clf.fig}b and c, we have also
indicated our conservative detection limit at $V = 22.5$ mag (for which
we are confident to have an almost fully complete sample; see de Grijs
et al.  [2001], their Fig.  7), or $M_V = -5.3$ mag (in case of no
extinction), at an age of 1.0 Gyr, assuming a distance to M82 of $m-M =
27.8$ (de Grijs et al.  2001).  Since {\it all} sources in our sample
are resolved, the sample is not contaminated by bright stars (see de
Grijs et al.  [2001] for a full discussion).  We have shown that if we
do not restrict ourselves to the limited age range of the burst, the age
and mass distributions obtained for the subsample with well-determined
ages and those for the full sample of M82 B clusters are internally
fully consistent (de Grijs et al.  2003).  However, if we impose age
limits on our analysis in order to restrict our study to a coeval
cluster population at $\sim 1$ Gyr, the large uncertainties in the age
determinations for the subset of the full sample with less
well-determined ages lead us to conclude that the results for the full
sample should be given less weight and be treated with caution.  Those
clusters that are not well fit (e.g., having less well-constrained ages
and masses) are generally fainter and as such are artificialy skewed
towards younger ages due to the well-known age-extinction degeneracy,
which is more important for these clusters owing to their lower
signal-to-noise ratios (see de Grijs et al. 2003 for more details). 

\begin{figure}
\plotone{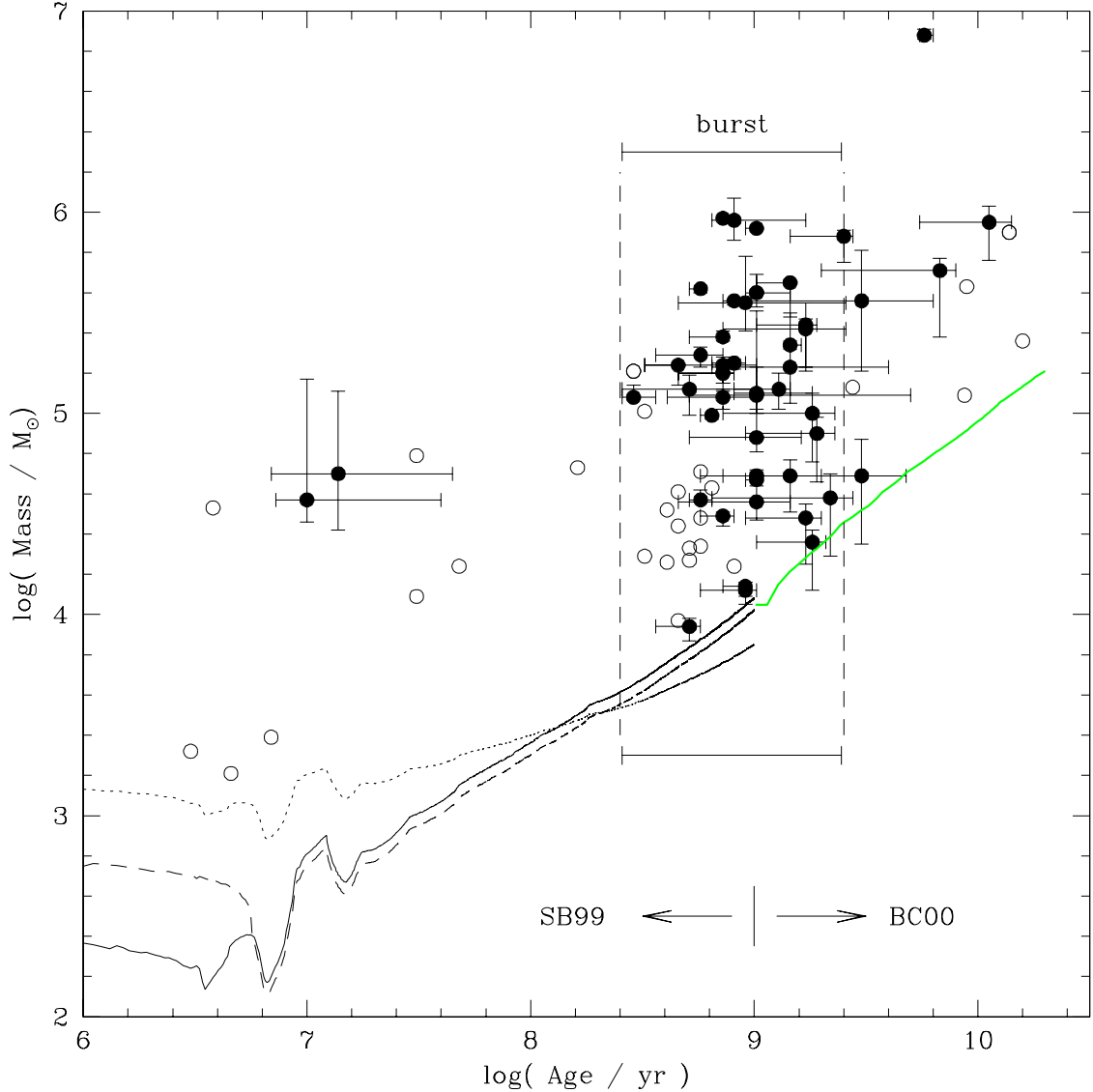}
\figcaption[f2.ps]{\label{m82agemass.fig} Distribution of the M82 B
clusters in the (age vs.  mass) plane.  The age range dominated by the
burst of cluster formation is indicated.  The black dots represent
clusters for which the total age range obtained is ($\log( {\rm
Age[max]} ) - \log( {\rm Age[min]} )) \le 1.0$; the open circles are
objects with more uncertain age determinations.  Overplotted for ages up
to 1.0 Gyr are the expected detection limits in the predicted by
Starburst99 (SB99) for a range of IMFs (solid line: mass range $0.1-100
M_\odot$, IMF slope $\alpha = 2.35$; dotted line: identical mass range,
but $\alpha = 3.30$; dashed line: mass range $0.1-30 M_\odot$, $\alpha =
2.35$); for older ages, we use the BC00 models for a standard IMF (mass
range $0.1-100 M_\odot$, $\alpha = 2.35$).  These model predictions are
based on a very conservative detection limit of $V = 22.5$ (see de Grijs
et al.  2001) and $(m-M)_{{\rm M}82} = 27.8$, assuming no extinction. 
For a nominal extinction of $A_V = 0.2$ mag, expected for the clusters
with well-determined ages (de Grijs et al.  2003), the detection limit
is expected to shift to higher masses by $\Delta \log( M_{\rm cl}/
M_\odot ) = 0.08$, which is well within the uncertainties associated
with our mass determinations.}
\end{figure}

In Fig.  \ref{m82agemass.fig} we show the distribution of the M82 B
clusters in the age vs.  mass plane.  The various (solid, dashed and
dotted) lines overplotted on the figure show the expected effect of
normal evolutionary fading of a synthetic single stellar population of
an instantaneously formed cluster at our limiting magnitude.  For ages
$\le 10^9$ yr, we show the unreddened fading line for various choices
for the IMF, predicted by the Starburst99 (SB99) models (Leitherer et
al.  1999).  For older ages ($t \ge 10^9$ yr), we show its extension
predicted by the BC00 models.  These predicted lower limits agree well
with our data points, which therefore shows that we understand our
selection effects to sufficient accuracy. 

\section{Detection of a turn-over in an intermediate-age CLF}

Both the M82 B burst CLF and the corresponding mass distribution show a
clear turn-over at about 2 magnitudes brighter, and an order of
magnitude more massive, than our detection limit, respectively.  This is
the first time that a turn-over has been detected for a coeval star
cluster system as young as $\sim 1$ Gyr. 

The characteristic turn-over mass of the M82 B clusters formed in the
burst of cluster formation, $M_{\rm TO} \simeq 1.2 \times 10^5 M_\odot$
(de Grijs et al.  2003), is approximately half that of the old Galactic
globular cluster system (Harris 1996).  However, if we assume a
power-law or Schechter-type initial cluster mass distribution in a Milky
Way-type gravitational potential with strongly radially dependent
radial anisotropy (Fall \& Zhang 2001), the peak of the mass
distribution will change over time towards higher masses due to the
effects of cluster disruption, which will preferentially deplete the
lower-mass clusters.  Furthermore, these models by Fall \& Zhang (2001)
suggest that the turn-over of the cluster mass distribution will move
towards higher masses by approximately $\Delta \log (M_{\rm cl} /
M_\odot) \simeq +0.9$ by the time the cluster population reaches an age
of 12 Gyr, similar to the median age of the Galactic globular cluster
system.  This implies that the star cluster system in M82 B will be
dominated by higher masses than the Galactic globular cluster system
when it reaches a similar age, and most of the present-day clusters will
be depleted.  This is most likely due to the fact that M82 B is
characterized by the shortest known cluster disruption time-scale for
any disk region of a galaxy, $\sim 30$ Myr for $10^4 M_\odot$ clusters
(de Grijs et al.  2003, using the method described in Boutloukos \&
Lamers 2002). 

This very short disruption time-scale suggests a significantly different
gravitational potential, however, so that our estimate of the amount
that the turn-over mass will move towards higher masses is, in fact, a
lower limit.  Additional support for this conclusion is given by the
observation that the current turn-over mass is already $\Delta \log
(M_{\rm cl} / M_\odot) \simeq +0.5$ more massive than the turn-over mass
for a young star cluster system in a Galactic gravitational potential at
a similar age, corresponding to $M_{\rm TO,young} \sim 7 \times 10^4
M_\odot$ (Fall \& Zhang 2001).\footnote{We note, however, that a
significant contributor to this apparent difference may be hidden in our
choice for the IMF, which we have assumed to be roughly Salpeter-like. 
Despite a growing body of evidence for a universal IMF independent of
the environment (see Gilmore [2001] for a review), significant
departures from a Salpeter-type IMF (e.g., Smith \& Gallagher 2001) may
give rise to larger-than-expected uncertainties in the derived mass
distribution.} In addition, the width of the mass distribution in Fig. 
\ref{m82clf.fig}c is significantly smaller than that of the Galactic
globular cluster mass function (Harris 1996), which again emphasizes the
significant effects of the very short cluster disruption time-scale and
the significantly different (and time-varying) gravitational potential
governing the M82 system. 

If the Fall \& Zhang (2001) models would apply to the M82 B cluster
system, which we assume for the sake of the current discussion, the peak
luminosity and width of the CLF will remain virtually unchanged for a
Hubble time from its current age of $\sim 1$ Gyr (S.M.  Fall, priv. 
comm.), due to the combination of evolutionary fading and continuing
disruption processes (Whitmore et al.  2002).  In that case, we can
directly compare the intermediate-age M82 B clusters to the old Galactic
globular cluster population.  The CLF of the M82 B cluster system is
characterized by a turn-over luminosity of $M_V^0 = -7.3 \pm 0.1$ mag,
and (within the Poissonian observational uncertainties) a Gaussian FWHM
of $\sim 3.1$ mag.  The M82 B CLF of clusters formed in the burst of
cluster formation is therefore nearly identical to the Galactic globular
CLF, although the old Galactic globular clusters are significantly more
metal-poor than the roughly solar-abundance M82 B clusters (Parmentier
et al.  2003).  This difference in metallicity will, however, only
affect the peak luminosity slightly, by less than $\sim 0.3$ mag
(Whitmore et al.  2002). 

Finally, we note that the key ingredient of the Fall \& Zhang (2001)
models is the radial anisotropy, which is a strong function of
galactocentric distance.  The net result of this assumption is that all
of the Galactic globular clusters in their models have effectively the
same pericenter.  While this does not have to apply to the cluster
system of M82 as a whole, we emphasize that the cluster system discussed
in this {\it letter} is confined to a relatively small spatially
confined region in the disk of M82, at the end of the central bar (cf. 
de Grijs 2001).  These clusters have, therefore, very similar
pericenters indeed. 

\section{Summary and Implications}

Thus, here we have presented the first conclusive evidence for a clear
turn-over in the CLF of a 1 Gyr-old, roughly coeval cluster population. 
The CLF shape and characteristic luminosity is nearly identical to that
of the apparently universal CLFs of the old globular cluster systems in
the Galaxy, M31, M87, and old elliptical galaxies (e.g., Whitmore et al. 
1995, Harris 1996, 2001, Ashman \& Zepf 1998, Harris et al.  1998). 
This is likely to remain virtually unchanged for a Hubble time.  We have
also shown that with the very short characteristic cluster disruption
time-scale governing M82 B, its cluster mass distribution will evolve
towards a higher characteristic mass scale than for the Galactic
globular clusters by the time it reaches a similar age.  We argue,
therefore, that this evidence, combined with the similar cluster sizes
(de Grijs et al.  2001), lends strong support to a scenario in which the
current M82 B cluster population will eventually evolve into a
significantly depleted old Milky Way-type globular cluster system
dominated by a small number of high-mass clusters.  This implies that
globular clusters, which were once thought to be the oldest building
blocks of galaxies, are still forming today in galaxy interactions and
mergers.  However, they will likely be more metal-rich than the
present-day old globular cluster systems. 

This connection between young or intermediate-age star cluster systems
and old globular clusters lends support to the hierarchical galaxy
formation scenario. Old globular clusters were once thought to have
been formed at the time of, or before, galaxy formation, i.e., during
the first galaxy mergers. However, here we have shown that the evolved
CLF of the compact star clusters in M82 B most likely to survive for a
Hubble time will probably resemble the high-mass wing of the
``universal'' old globular cluster systems in the local Universe. 
Proto-globular cluster formation thus appears to be continuing until the
present. 

In order to better constrain the future evolution of the M82 B star
cluster system it is important to consider a range of models,
characterized by fewer or perhaps different orbital restrictions (e.g.,
Baumgardt 1998, Vesperini 2000, 2001).  This is, however, beyond the
scope of the current {\it letter} and will be done in a subsequent paper
(de Grijs et al., in prep.). 

\paragraph{Acknowledgements} - We acknowledge useful and interesting
discussions with R.W.  O'Connell and S.M.  Fall.  We also thank A. 
Helmi for a critical reading of the manuscript, and the anonymous
referee for helpful suggestions.  RdeG acknowledges support from the
Particle Physics and Astronomy Research Council (PPARC) and from The
British Council under the {\sl UK--Netherlands Partnership Programme in
Science}.

\end{document}